\documentclass[twocolumn,showpacs,preprintnumbers,amsmath,amssymb]{revtex4}
\input psfig.sty

\usepackage{graphicx}
\usepackage{dcolumn}
\usepackage{bm}

\begin{document}

\def\be{\begin{equation}}
\def\ee{\end{equation}}


\title{Observational constraints on the time-dependence of dark energy}

\author{Deepak Jain} \email{deepak@physics.du.ac.in}
\affiliation{Deen Dayal Upadhyaya College, University of Delhi, Delhi - 110015, India
}

\author{J. S. Alcaniz} \email{alcaniz@on.br}

\affiliation{Observat\'orio Nacional, Rua Gal. Jos\'e Cristino 77, 20921-400 Rio de Janeiro - RJ,
Brasil}

\author{Abha Dev} 
\affiliation{Department of Physics and Astrophysics, University of Delhi, Delhi -
110007, India
}

\date{\today}

\begin{abstract}

One of the most important questions nowadays in physics concerns the nature of the
so-called dark energy. It is also a consensus among cosmologists that such a question
will not be answered on the basis only of observational data. However, it is possible
to diminish the range of possibilities for this dark component by comparing different
dark energy scenarios and finding which models can be ruled out by current
observations. In this paper, by asssuming three distinct parametrizations for the
low-redshift evolution of the dark energy equation of state (EOS), we consider the
possibility of discriminating between evolving dark energy and $\Lambda$CDM models from
a joint analysis involving the most recent radio sources gravitational
lensing sample, namely, the Cosmic All Sky Survey (CLASS) statistical data and the
recently published \emph{gold} SNe Ia sample. It is shown that
this particular combination of observational data restricts considerably the dark
energy parameter space, which enables possible distinctions between time-dependent and
constant EOS's.

\end{abstract}

\maketitle

\section{Introduction}

The idea of a negative-pressure dark component which accounts for $\sim 2/3$ of the
critical density seems to be strongly supported by the current cosmological
observations. 
Very little, however, is known about the nature of this extra component, a fact that has
opened the possibility for many speculations on its fundamental origin and has also
given rise to the so-called dark energy
problem. These speculations are usually based either on a particular choice of the
equation of state (EOS) characterizing the dark energy ($\omega \equiv p/\rho$)
\cite{eos} or on
modifications of gravity at very
large scales \cite{brane}. However, irrespective of the choice made, a consensus is
being
reached nowadays on the fact that is extremely difficult, if not impossible, to single
out the best-fit model of dark energy on the basis only of observational data (see
\cite{1} for some attempts to physically constrain the nature of the dark energy). 

The above  conclusion, in turn, leads inevitably to a pessimist attitute towards the
observational investigation of
dark energy. However, as well observed by Kratochvil {\it et al.} \cite{krat}, one may,
instead  of trying to find which dark energy scenario is correct, test which
models can be ruled out by the available observations. This, of course, will not unveil
the nature of the dark energy but may, with the current and upcoming observational data,
diminish considerably the range of possibilitites. An interesting example
involves two of the favorite candidates for dark energy, namely, the vacuum energy
density or the  cosmological constant ($\Lambda$) and a dynamical scalar field
($\phi$), usually called \emph{quintessence}. Among other things, what observationally
differs these two candidates for dark energy is that, in the former case, the EOS 
associated with $\Lambda$ is constant along the evolution of the Universe
($\omega = -1$) whereas in generic quintessence scenarios $\omega$ is a function of
the scalar field $\phi(t)$ as well as  of its potential $V(\phi)$ \footnote{A time-dependent EOS is also a feature of some braneworld scenarios. See, e.g., \cite{deff}.}. Therefore, taking this small but important difference into account, one
may conclude that if any observable deviation from a constant equation of state is
consistently found, this naturally poses a problem for any model based on this
assumption, which includes our current concordance scenario ($\Lambda$CDM). 

Following this reasoning, we aim, in this paper, to explore the prospects for
constraining possible time
dependence of dark energy from a joint analysis involving radio-selected gravitational
lens statistics and
supernova (SNe Ia) data. To this end, we use the most recent radio sources gravitational
lensing sample, namely, the Cosmic All Sky Survey (CLASS) statistical data which
consist of 8958 radio sources out of which 13 sources are multiply imaged
\cite{chae} and the recently published SNe Ia data set with a total of 157 events
\cite{rnew}. This particular
combination of lensing statistics and SNe Ia data has been used by some authors to this
very same end \cite{huterer} (and also to place limits on constant EOS models 
\cite{waga,dev1}) and constitutes a potential probe for 
possible variations of the dark energy EOS since it covers a considerable
interval of redshift, which is a necessary condition to properly distinguish
redshift-dependent equation of states from models with constant $\omega$. By considering
three 
different parametrizations for the time-dependence of the dark energy, it is shown that
this combination of observational data restricts considerably the dark
energy parameter space, which enables possible distinctions between time-dependent and
constant EOS's.

\section{$\omega(z)$ Models}

In this work, we are particularly interested in three specific parametrizations for the
variation of $\omega$ with redshift, i.e.,
\begin{equation}
\omega(z) = \omega_o + \omega_1 z   \quad    \quad    \quad   \quad    \quad   \quad   
\quad          
\quad (\mbox{P1}),
\end{equation}
\begin{equation}
\omega(z) = \omega_o - \omega_2{\rm{ln}}(1 + z)\quad\quad \quad  \quad \quad 
(\mbox{P2}),
\end{equation}
and
\begin{equation}
\omega(z) = \omega_o +  \omega_3(\frac{z}{1 + z}) \quad  \quad     \quad  \quad    \quad
(\mbox{P3})
\end{equation}
where $\omega_o$ is the current value of the equation-of-state parameter, and $\omega_j$
($j = 1, 2, 3$) are free parameters quantifying the time-dependence of the dark energy
EOS, 
which must be adjusted by the observational data. Note that the EOS
of the cosmological constant can be always recovered by taking $\omega_j = 0$ and
$\omega_o = -1$.

The Taylor expansion (P1) was firstly suggested in Ref. \cite{p1}. Constraints on (P1)
were firstly studied by Cooray \& Huterer
\cite{huterer} by using SNe Ia data, gravitational
lensing statistics and globular clusters ages and also by Goliath {\it et al.}
\cite{gol} who investigated limits to this parametrization from future SNe Ia
experiments. As commented in Ref. \cite{huterer},
P1 is a good approximation for most quintessence models out to redshift of a few and it
is exact for models where the equation of state is a constant or changing slowly. P1,
however, has serious problems to explain age estimates of high-$z$ objects since it
predictes very small ages at $z \geq 3$ \cite{friaca} (In reality, P1 blows up at
high-redshifts as $e^{3\omega_1z}$ for values of $\omega_1 > 0$ -- see Eq. (4) below).
The
empirical fit P2 was introduced by Efstathiou \cite{efs} who argued that for a wide
class of potentials associated to dynamical scalar field models the evolution of
$\omega(z)$ at $z \lesssim 4$ is well approximated by Eq. (2). P3 was recently proposed
in Refs. \cite{chev,linder} (see also \cite{pad}) aiming at solving undesirable
behaviours of P1 at high redshifts. According to \cite{krat}, this parametrization is a
good fit for many theoretically conceivable scalar field potentials, as well as for
small recent deviations from a pure cosmological constant behaviour ($\omega = -1$) [see
also \cite{wsps,teg,padn} for other parametrizations].

Since Eqs. (1-3) represent separately conserved components, it is straightforward to
show from the energy conservation law [$\dot{\rho_j} = -3\dot{R}/R(\rho_j + p_j)$] that
the ratio ${f_j} = \rho_j/{\rho_j}_o$ for (P1)-(P3) evolves, respectively, as
\begin{equation}
{f_1} = (\frac{R_o}{R})^{3(1 + \omega_o - \omega_1)}\rm{exp}\left[3 
\omega_1 (\frac{R_o}{R} - 1)\right] ,
\end{equation}
\begin{equation}
f_2 = (\frac{R_o}{R})^{3\left[1 + \omega_o
-\frac{\omega_2}{2}\rm{ln}(\frac{R_o}{R})\right]} ,
\end{equation}
\begin{equation}
{f_3} = (\frac{R_o}{R})^{3(1 + \omega_o +
\omega_3)}\rm{exp}\left[3\omega_3(\frac{R}{R_o} - 1)\right],
\end{equation}
where the subscript $o$ denotes present day quantities and $R(t)$ is the cosmological
scale factor. The distance-redshift and the age-redshift relations -- two fundamental
quantities related to
the observables which will be considered in the next section -- are given respectively
by
\begin{equation} 
\xi_j(z) = \frac{c}{R_o H_o}\int_{o}^{z} {dz \over  \left[\Omega_{\rm{m}}(1 + z)^{3} +
(1 - \Omega_{\rm{m}}) f_j\right]^{1/2}} 
\end{equation} 
and
\begin{equation} 
\tau_j(z) = \frac{1}{H_o}\int_{o}^{z} {(1 + z)^{-1} dz \over  \left[\Omega_{\rm{m}}(1 +
z)^{3} +
(1 - \Omega_{\rm{m}}) f_j\right]^{1/2}}, 
\end{equation}
where $\Omega_{\rm{m}}$ stands for the matter density parameter. Throughout this paper we fix
$\Omega_{\rm{m}} = 0.3$, in accordance with several dynamical estimates of the quantity
of matter in the Universe \cite{calb}.

\section{Lensing and SNe Ia Constraints}

In our search to constrain a possible time-dependence of the dark energy, we adopt a
joint analysis involving the so far
largest lensing sample suitable for statistical analysis along with the latest SNe Ia data,
as provided by Riess {\it et al.} \cite{rnew}. In what follows, we discuss both the lensing and SNe Ia samples used as well as the main assumptions on which we performed our joint analysis.

\subsection{CLASS statistical Sample} 

The final CLASS well-defined statistical sample consists of 8958 radio sources out of which 13 sources are multiply imaged. Here we work only with those multiply imaged sources whose image-splittings are known (or likely) to be caused by single galaxies. There are 9 such radio sources: 0218+357, 0445+123, 0631+519, 0712+472, 0850+054, 1152+199, 1422+231, 1933+503, 2319+051. We thus work with a total of 8954 radio sources \cite{chae1,dev}. The sources probed by CLASS at $\nu = 5$ GHz are well represented by power-law differential number-flux density relation: $\left |dN/dS \right| \propto (S/S^{0})^{\eta}$ with $\eta = 2.07 \pm 0.02$ ($1.97 \pm 0.14$) for $S \geq S^{0}$ ($ \leq S^{0}$) where $S^{0} = 30$ mJy \cite{bro}. The CLASS unlensed sources can be adequately described by a Gaussian model with  mean redshift, $z = 1.27$ and a dispersion of $0.95$.

We start our analysis by assuming the singular isothermal sphere
(SIS) model for the lens mass distribution. As has been discussed
elsewhere this assumption represents a good approximation to the
real mass distribution in galaxies (see, e.g., \cite{TOG}). For
the present analysis we also ignore the evolution of the number
density of galaxies and assume that the comoving number density is
conserved. The present day comoving number density of galaxies is
\begin{equation} \label{number}
n_o = \int_0^{\infty} \phi(L) dL,
\end{equation}
where  $\phi (L)$ is the well known Schechter Luminosity Function
\cite{sch}.

The differential optical depth of lensing in traversing $dz_L$
with angular separation between $\phi$ and $\phi + d\phi$ is
\cite{glambda}:
\begin{eqnarray}
\frac{d^{2}\tau}{dz_{L}d\phi}d\phi dz_{L} &=& F^{*}\,(1 +
z_{L})^{3}\,\left({{D_{OL} D_{LS}}\over{ R_{o}
D_{OS}}}\right)^{2}\ \frac{1}{R_{o}}\, \frac{cdt}{dz_{L}}
\nonumber \\ && \times
\frac{\gamma/2}{\Gamma(\alpha+1+\frac{4}{\gamma})}
\left(\frac{D_{OS}}{D_{LS}}\phi\right)^{\frac{\gamma}{2}(\alpha+1+\frac{4}{
\gamma})} \nonumber \\ && \times
{\rm{exp}}\left[-\left(\frac{D_{OS}}{D_{LS}}\phi\right)^{\frac{\gamma}{2}
}\right]\frac{d\phi}{\phi} dz_{L}, \label{diff}
\end{eqnarray}
where the function $F^{*}$ is defined as
\begin{equation}
F^* = {16\pi^{3}\over{c\, H_{0}^{3}}}\phi_\ast
v_\ast^{4}\Gamma\left(\alpha +{4\over\gamma} +1\right).
\end{equation}
The quantities $D_{OL}$, $D_{OS}$ and $D_{LS}$ represent, respectively, the angular diameter distances from the observer to the lens, from the observer to the source and between the lens and the source. In order to relate the characteristic luminosity $L_*$ to the characteristic velocity dispersion $v_{*}$, we use the Faber-Jackson relation \cite{fj} for early-type galaxies ($L_*
\propto {v_*}^{\gamma}$), with $\gamma = 4$. For the analysis presented here we neglect the contribution of spirals as lenses because their velocity dispersion is small when compared to
ellipticals.

\begin{figure*}
\vspace{.2in}
\centerline{\psfig{figure=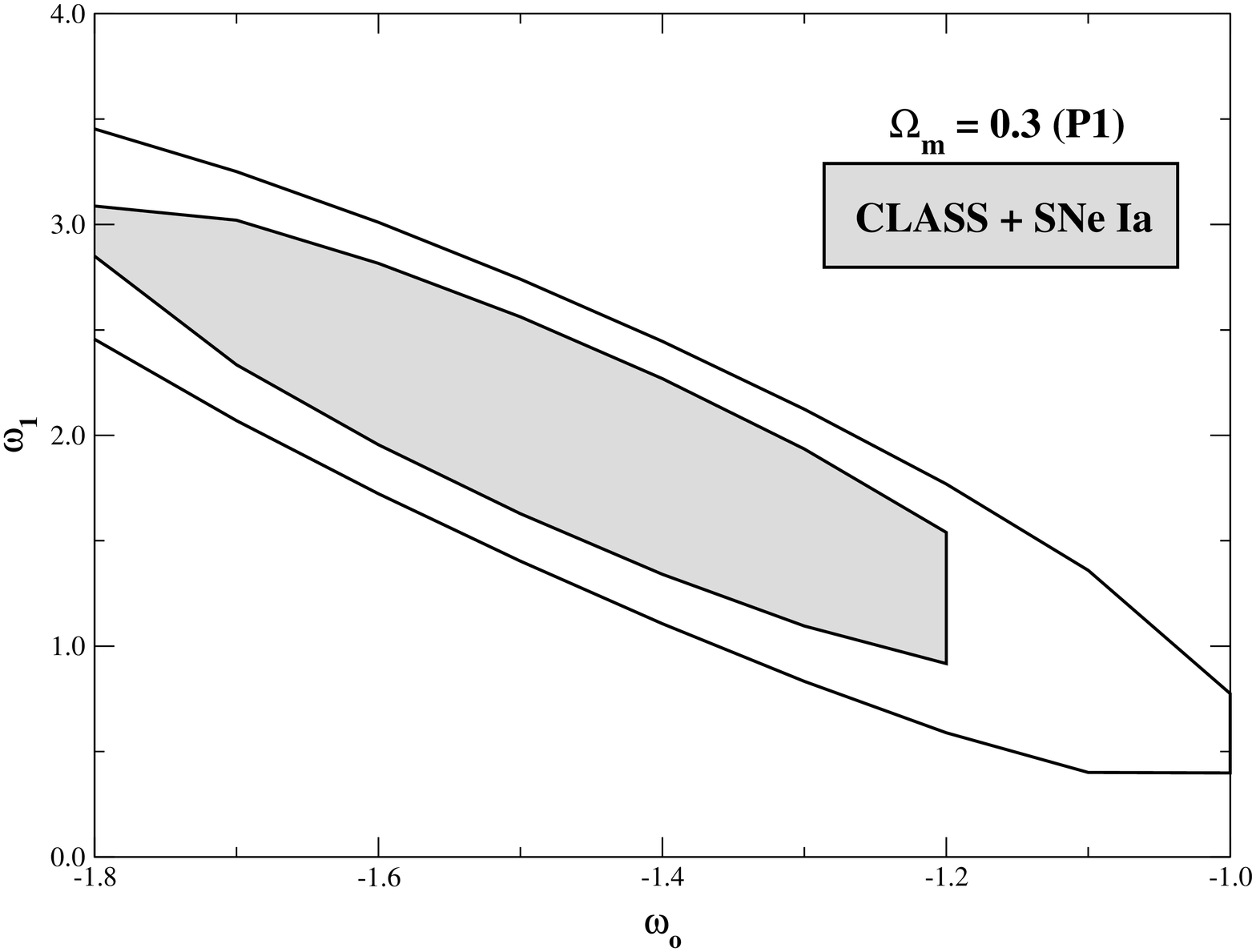,width=2.4truein,height=3.3truein,angle=-90} 
\psfig{figure=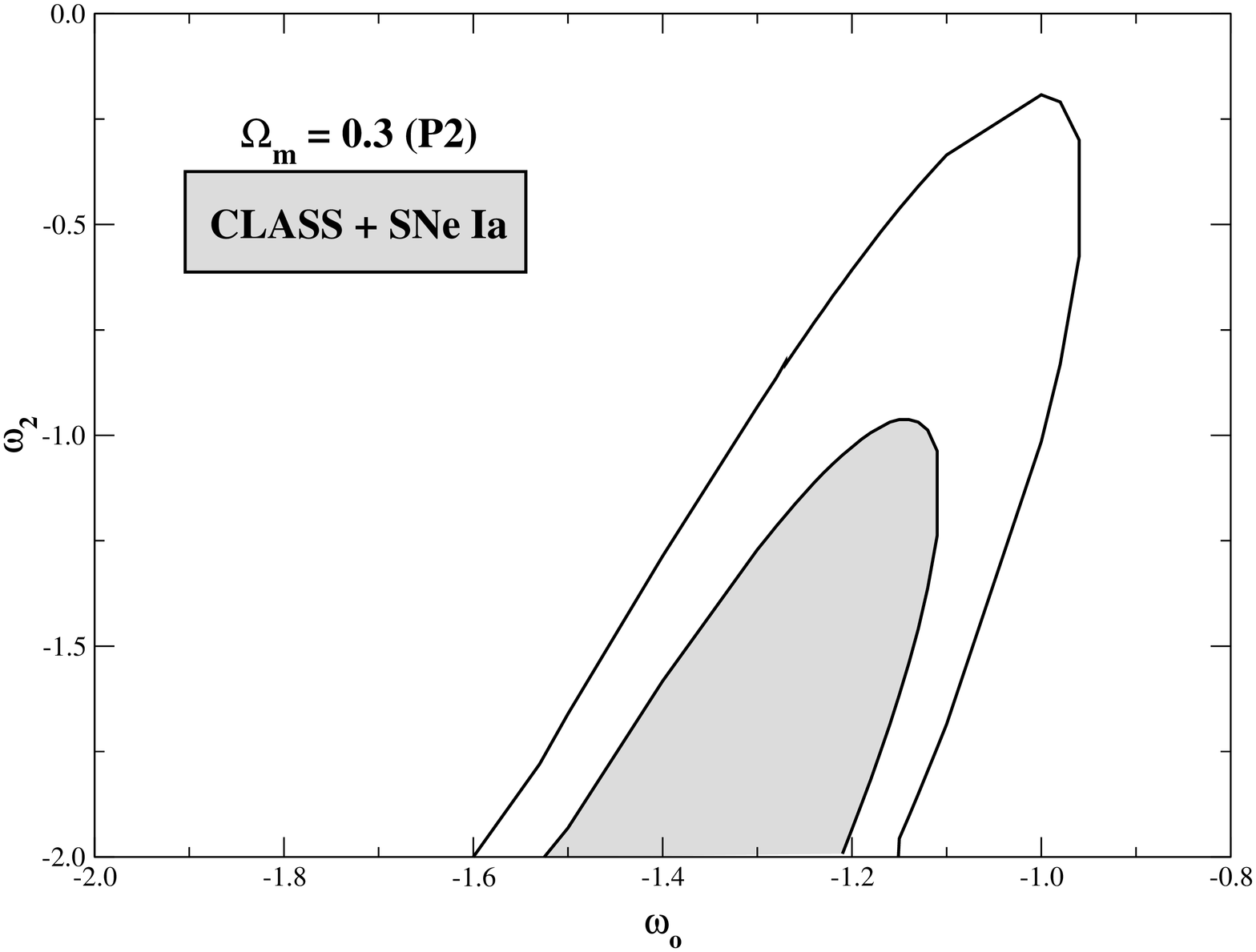,width=2.4truein,height=3.3truein,angle=-90}
\psfig{figure=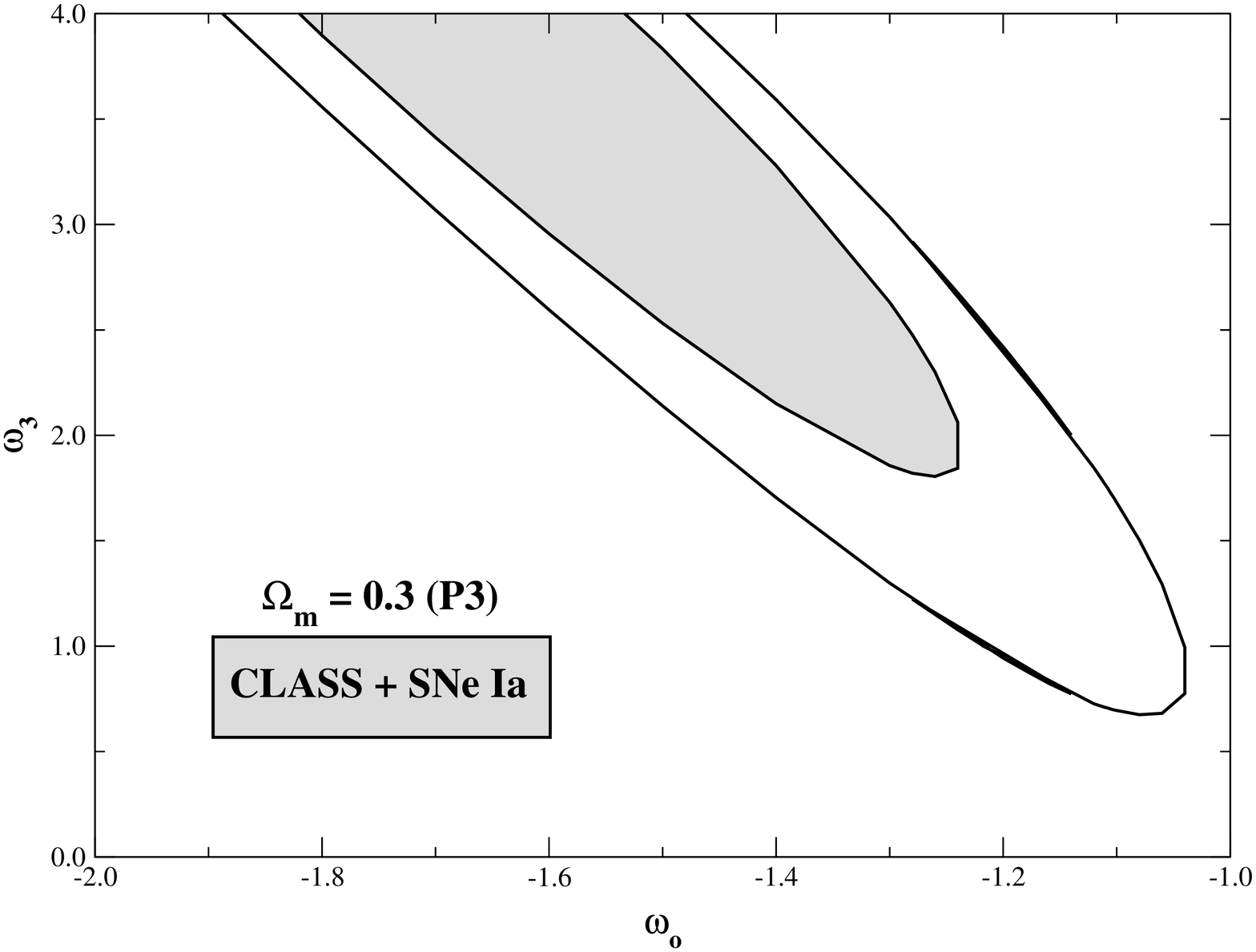,width=2.4truein,height=3.3truein,angle=-90}
\hskip 0.1in}
\caption{Current constraints on the plane $\omega - \omega_j$ from CLASS lensing
statistics and SNe Ia data. Note that for the three different parametrizations
considered in this paper the joint CLASS + SNe Ia analysis
clearly prefers the regions where $\omega_j \neq 0$, i.e., a time-dependent EOS.
{\bf{a)}} Confidence regions (68.3\% and 95.4\%) in the plane $\omega - \omega_1$ for
P1. {\bf{b)}} The same as in Panel 1a for P2. {\bf{c)}} The same as in Panel 1a for
P3.}
\end{figure*}

The two large-scale galaxy surveys, namely, the 2dFGRS \footnote{The 2dF Galaxy Redshift-Survey (2dfGRS): http://msowww.anu.edu.au/2dFGRS/} and the SDSS \footnote{Sloan Digital Sky Survey: http://www.sdss.org/} have produced converging results on the total LF. The surveys determined the Schechter parameters for galaxies (all types) at $z \le 0.2$. Chae \cite{chae1} has worked extensively on the information provided by these recent galaxy surveys to extract the local type-specific LFs. For our analysis here, we adopt the normalization corrected Schechter parameters of the 2dFGRS survey \cite{chae1,folkes}: $\alpha = -0.74$, $\phi^{*} = 0.82 \times 10^{-2} h^3
\mathrm{Mpc^{-3}}$, $v^{*} = 185\,\mathrm{km/s}$ and $F^{*} = 0.014$.

The normalized image angular separation distribution for  a source at $z_{S}$ is obtained by integrating $\frac{d^{2}\tau}{dz_{L}\,d\phi}$ over $z_{L}$:

\begin{equation}
{d{\mathcal{P}}\over d\phi}\, =\,
{1\over\tau(z_S)}\int_{0}^{z_{s}}\,{\frac{d^{2} \tau
}{dz_{L}d\phi}} {dz_{L}}.\end{equation}

The corrected (for magnification and selection effects) image separation distribution function for a single source at redshift $z_{S}$ is given by \cite{CSK1}
 \begin{eqnarray}
 P'(\Delta\theta)\,& =& \, \mathcal{B}\,{{{\gamma}} \over
 2\, \Delta \theta} \int_{0}^{z_S}\left [{D_{OS}\over{D_{LS}}} \phi
 \right ]^{{\frac{\gamma}{2}}(\alpha + 1+ {\frac{4}{\gamma}})}
 \,F^{*}\,{cdt\over dz_{L}} \nonumber \\ && \times \exp\left[- \,\left({D_{OS}\over{D_{LS}}}
\phi\right)^{\frac{\gamma}{2}}\right] {(1 + z_{L})^{3} \over
\Gamma\left(\alpha +{4\over\gamma} +1\right)}  \nonumber \\ &&
\times \left[\,\left ({D_{OL}D_{LS}\over R_0
  D_{OS}}\right)^{2}\,{1\over R_0}\right ]\,\,{dz_{L}}.
\label{dist}
\end{eqnarray}

\noindent Similarly, the corrected lensing probability for a given source at redshift $z$ is given  by
\begin{equation}
P' = \tau(z_S) \,\int {d{\mathcal{P}}\over d\phi} \mathcal{B}\;
d\phi. \label{prob}
\end{equation}
Here $\phi$ and $\Delta\theta$ are related to as $\phi = {\frac{\Delta\theta}{8 \pi (v^{*}/c)^2}}$ and $\mathcal{B}$ is the magnification bias. This is considered because, as is widely known,
gravitational lensing causes a magnification of images and this transfers the  lensed sources to higher flux density bins. In other words, the lensed sources are over-represented in a
flux-limited sample. The magnification bias  ${\mathcal B}(z_S,S_\nu)$ increases the lensing probability significantly in a bin of total flux density ($S_\nu$) by a factor
 \begin{eqnarray}
\mathcal{B}(z_S,S_\nu) \,& =& \,  \left
|\frac{dN_{z_S}(>S_\nu)}{dS_\nu} \right|^{-1} \\ && \times
\int_{\mu_{min}}^{\mu_{max}} \left
|\frac{dN_{z_S}(>S_\nu/\mu)}{dS_\nu}\,p(\mu)\right |{1\over\mu}
\,d\mu.  \nonumber
\label{B2}
\end{eqnarray}
In the above expression, $N_{z_{S}}(> S_\nu)$ is the intrinsic flux density relation for the source population at redshift $z_{S}$. $N_{z_{S}}(> S_\nu)$ gives the number of sources at redshift $z_{S}$ having
flux greater than $S_\nu$. For the SIS model, the magnification probability distribution is $p(\mu) = 8/{\mu}^{3}$. The minimum and maximum total magnifications $\mu_{min}$ and $\mu_{max}$ in
equation (\ref{B2}) depend on the observational characteristics as well as on the lens model. For the SIS model, the minimum total magnification is $\mu_{min} \simeq 2$ and the maximum total
magnification is $\mu_{max} = \infty$. The magnification bias $\mathcal B$ depends on the differential number-flux density relation $\left|dN_{z_{S}}(> S_{\nu})/dS_{\nu}\right|$. The
differential number-flux relation needs to be known as a function of the source redshift. At present redshifts of only a few CLASS sources are known. We, therefore, ignore redshift dependence of the differential number-flux density relation. Following Chae \cite{chae1}, we further ignore the dependence of the differential number-flux density relation on the spectral index of
the source.

Two important selection criteria for CLASS statistical sample
are (i) that the ratio of the flux densities of the fainter to the
brighter images ${\mathcal R}_{min}$ is $ \ge 0.1$. Given such an
observational limit, the minimum total magnification for double
imaging the adopted model of the lens is $\mu_{min} = 2 (1+\mathcal{R}_{min}/1-\mathcal{R}_{min})$ \cite{chae1}; (ii) that the image components in lens
systems must have separations $\ge 0.3$~arcsec. We incorporate
this selection criterion by setting the lower limit of $\Delta
\theta$ in equation (\ref{prob}) as $0.3$ arcsec.

\subsection{SNe Ia sample.} 

The SNe Ia sample of Riess {\it et al.} \cite{rnew} consists of
186 events distributed
over the redshift interval $0.01
\lesssim z \lesssim 1.7$ and constitutes the compilation of the best observations made
so far by the two supernova search teams plus 16 new events observed by \emph{HST}.
This total data-set was divided into ``high-confidence'' (\emph{gold}) and ``likely but
not certain'' (\emph{silver}) subsets.  Here, we will consider only the 157 events that
constitute the so-called \emph{gold} sample.  The best fit to the set of parameters
$\mathbf{s}$ is obtained by using a $\chi^{2}$ statistics, i.e., 
\begin{equation}
\chi^{2}_{SNe} =
\sum_{i=1}^{157}{\frac{\left[\mu_p(z|\mathbf{s}) -
\mu_o^{i}(z|\mathbf{s})\right]^{2}}{\sigma_i^{2}}},
\end{equation} 
where $\mu_p(z|\mathbf{s}) = m - M = 5\mbox{log} d_L + 25$ is the predicted distance
modulus for a supernova at
redshift $z$, $\mu_o^{i}(z|\mathbf{s})$ is the
extinction corrected distance modulus for a given SNe Ia at $z_i$, and $\sigma_i$ is
the uncertainty in the individual distance moduli, which includes uncertainties in
galaxy redshift due to a peculiar velocity of 400 km/s. The Hubble parameter $H_o$ is
considered a ``nuisance" parameter so that we marginalize over it (For some recent SNe Ia studies, see \cite{sne}).

\subsection{Results.} 

The main results of our joint analysis are displayed in Figure 1. Panels
1a-1c show the
confidence regions ($68.3\%$ and $95.4\%$) in the $\omega_{o} - \omega_{j}$ plane for
P1, P2 and P3, respectively. As stated earlier, the matter density parameter has been
fixed in all analyses at $\Omega_m = 0.3$, in agreement with some clustering estimates
\cite{calb}. The contours are defined by the
conventional two-parameter
$\chi^2$ levels (2.30 and 6.17), where $\chi^{2}_{total} = \chi^{2}_{SNe} -
2{\rm{ln}}l$ and $l = {\cal{L}}_{lens}/{\cal{L}}_{max}^{lens}$ is the normalized
likelihood for lenses. From this combination of observational data we find that the
best-fit parameters for P1 are $\omega_o = -1.5$ and $\omega_1 = 2.1$, which
corresponds to an accelerating scenario with transition redshift (at which the Universe
switches from acceleration to acceleration) $z_t \simeq 0.26$ and a total expanding age
of $\simeq 7.0h^{-1}$ Gyr. For parametrizations 2 and 3 the best-fit scenarios occur at
$\omega_o = -1.4$ and $\omega_2 = -2$ and $\omega_o = -1.7$ and $\omega_3 = 4.1$,
corresponding to a $10h^{-1}$-Gyr-old and $7.5h^{-1}$-Gyr-old universes with transition
redshifts at $z_t \simeq 0.53$ and $z_t \simeq 0.31$, respectively. Note that the
estimates of $z_t$ for P2 and P3 are inside the 1$\sigma$ interval inferred for the
transition redshift from the current SNe Ia data \cite{rnew} (see also \cite{daly}). The
best-fit values for each parametrization also imply a beginning of a phantom behavior
at $z_{ph} \simeq 0.23$ (P1), $z_{ph} \simeq 0.22$ (P2) and $z_{ph} \simeq 0.20$ (P3).

From the above figures, however, it is clear the most important conclusion one may reach
resides on
the fact that for 
the three parametrizations considered in this paper the joint CLASS + SNe Ia analysis
clearly 
prefers the regions where $\omega_j \neq 0$, i.e., a time-dependent EOS. In particular, 
note that the point $\omega_j = 0$, which means a time-independent EOS, is at least
2$\sigma$ off from the central values obtained for this parameter. Such a result seems
to be more restrictive than that obtained in Ref. \cite{teg} (see also \cite{padn}), in
which a combination of
the latest SNe Ia, cosmic microwave background and large-scale structure data showed no
hint of departures from the model corresponding to Einstein's
cosmological
constant ($\omega_j = 0$ and $\omega_o = -1$). It is possible that this particular
preference for a time-dependent EOS changes if the current cosmic microwave background
(CMB) data
were included in the analysis. As discussed in \cite{padn}, an evolving dark energy EOS
affects the features of temperature anisotropies in CMB in at least two ways, namely,
the position of the acoustic peaks as well as the integrated Sachs-Wolf efect. 
However, if independent analyses involving a more significant number of different data
sets confirm this preference of
the observational data for values of $\omega_j \neq 0$, these results surely will bring
to light a new consistency problem for our current standard cosmological model since for this case 
$\omega_j$ is necessarily null.

\section{Conclusion}

Very recently, the field of cosmology has entered a golden era. An era where new and
revolutionary concepts were introduced with the support of a plethora of high-quality
observational data. Surely, the most remarkable among these concepts is the idea of a
dark energy-dominated universe, which is motivated from an impressive
convergence of independent observational results. This idea in turn gave rise to the
so-called dark energy problem since the nature of this dark component is completely
unknown at present. 

In this paper, although aware of the impossibility of determining the nature of the dark
energy on the basis only of observational data, we have considered the possibility of 
discriminating two of our favorite candidates for this mysterious component, namely,
the cosmological constant ($\Lambda$) and a dynamical scalar field ($\phi$) [33]. By
considering three different parametrizations for the dark enegy EOS, we have placed
limits on the time-dependent term of these parametrizations ($\omega_j$) from a joint
analysis involving the most recent radio sources gravitational lensing sample and SNe
Ia data. We have shown that this particular combination of observational data prefer values of $\omega_j \neq 0$, i.e., a time-dependent EOS. We believe that if such
a result is confirmed by the upcoming observations, it may shed some light on our
search for a better understanding of the nature of the so-called dark energy.

\begin{acknowledgments}
The authors are very grateful to C. S. Vilar and P. Mehta for valuable discussions and a
critical reading of the manuscript. JSA is supported by Conselho Nacional de Desenvolvimento Cient\'{\i}fico e Tecnol\'ogico (CNPq/307860/2004-3).
\end{acknowledgments}


\end{document}